\documentclass[reprint,pre,twocolumn,amsmath,amssymb,longbibliography]{revtex4-1}
\usepackage{graphicx}
\usepackage{dcolumn}
\usepackage{bm}
\usepackage{epstopdf}
\usepackage{caption}
\usepackage{subcaption}

\newcommand{\be}{\begin{equation}}
\newcommand{\ee}{\end{equation}}
\newcommand{\beal}{\begin{align}}
\newcommand{\eal}{\end{align}}

\begin{document}

\title{Multifractal to monofractal evolution of the London's street network}
\author{Roberto Murcio}
\email{r.murcio@ucl.ac.uk}
\author{A. Paolo Masucci}
\email{a.masucci@ucl.ac.uk}
\author{Elsa Arcaute}
\email{e.arcaute@ucl.ac.uk}
\author{Michael Batty}
\email{m.batty@ucl.ac.uk}
\affiliation{
 Centre for Advanced Spatial Analysis. University College London \\1st Floor, 90 Tottenham Court Road London, United Kingdom 
}

\date{\today}

\begin{abstract}

We perform a multifractal analysis of the evolution of London's street network from 1786 to 2010. First, we show that a single fractal dimension, commonly associated with the morphological description of cities, does not suffice to capture the dynamics of the system. Instead, for a proper characterization of such a dynamics,  the multifractal spectrum needs to be considered. Our analysis reveals that London evolves from an inhomogeneous fractal structure, that can be described in terms of a multifractal, to a homogeneous one, that converges to monofractality. We argue that London's multifractal to monofracal evolution might be a special outcome of the constraint imposed on its growth by a green belt.
Through a series of simulations, we show that multifractal objects, constructed through diffusion limited aggregation, evolve towards monofractality if their growth is constrained by a non-permeable boundary.

\end{abstract}


\maketitle

\section{\label{sec:level1}Introduction}

Street networks are ubiquitous worldwide, forming possibly the most important and articulated infrastructural network. 
These have been considered to be mathematical objects since the eighteenth century \cite{Euler}, and with the discovery of Zipf's law and the conjecture of Gibrat's law \cite{Zipf,eeckhout2004gibrat,10.1257/aer.101.5.2205,gibrat1930loi} they have gained notable relevance within statistical physics.
In addition, urban patterns share many statistical and morphological similarities with biological and physical phenomena, suggesting that common organisational principles might be underlying these processes \cite{barenblatt1996scaling,batty2007cities}.

Such patterns can be analysed through fractal geometry. Originally formalized in the 70's by Mandelbrot \cite{mandelbrot1983fractal}, it wasn't until two decades later that fractals were applied to describe urban morphology \cite{battylf1996preliminary,frankhauser1998fractal,batty1994fractal,makse1995modelling} and urban growth \cite{makse1998modeling,rozenfeld2008laws}. 
Until recently, a single fractal dimension was employed. It is well recognised now, that a spectrum of fractal dimensions needs to be employed to fully characterise systems that present different fractal properties at different scales and regions \cite{stanley1988multifractal}, as is the case of urban systems \cite{appleby,ArizaVillaverde20131,Hu2012161}. These systems are called \emph{multifractals}.

Fractal objects are in general described through a single measure called the \textit{capacity} or \textit{box-counting} dimension, which measures the amount of space filled by the fractal, disregarding local density differences. 
These differences are nevertheless an important aspect to evaluate and compare systems according to their local characteristics, such as measuring performance in cities.
In addition, the box-counting dimension is extremely sensitive to external parameters, and has been proven to lead to poor results \cite{falconer2004fractal,Reeve:1992:WSE:149805.149814}.

A measure conveying more information about the internal components of a fractal structure, directly relating the local density to a measure of proximity between any pair of elements of the structure, is the \textit{correlation dimension} \cite{Theiler:90}. 
 
\begin{figure}[ht]
\includegraphics[width=1\linewidth]{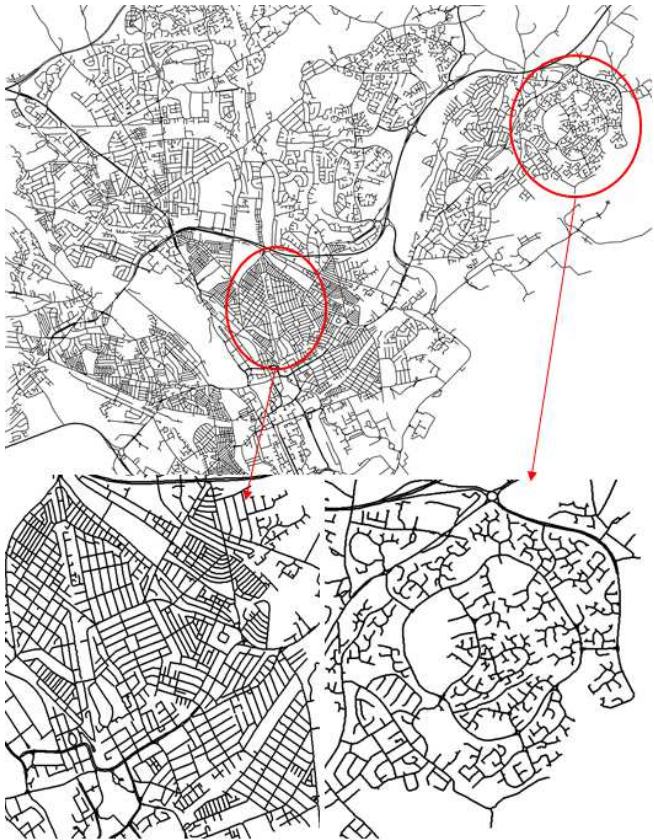}
\caption{(Color online) The street network of Cardiff, UK, along with a zoom of two particular areas.}
\label{I1}
\end{figure}

Nevertheless, this dimension only gives a general balance of the distribution of its elements, loosing any information on the heterogeneities related to particular zones.
This is illustrated in Fig.~\ref{I1}, where two very different areas are highlighted in the street network of Cardiff.

The single correlation dimension associated to the street network does not give any information on the differences between the north and the south layout. The latter has more regular and compact structures, while the north one tends to be more curly, with a minor number of intersections. 
These differences in density convey important information about the activities taking place in cities. 

In general, high densities of intersections are found only in a few places in a city, whereas the majority of the network has a small number of intersections.
Let us denote by $\alpha_0$ the areas with an intersection density larger than $\rho_0$, and by $\alpha_1$ the areas with an intersection density $\rho_1$ such that $\rho_1 < \rho_0$. The correlation dimension for each of the two sets given by $\alpha_0$ and $\alpha_1$ will be different. 
It is expected that the fractal dimension for the denser set ($\alpha_0$) will be close to two, while that for the set ($\alpha_1$) where the points are scattered all over the area, will be close to zero \cite{mandelbrot1983fractal}. 

There are many systems that behave in such a way, where a single fractal exponent is unable to capture the complexity of the fractal structure.
This is the basis for \textit{multifractal analysis}, which was introduced by Maldelbrot to study turbulent flows \cite{FLM:385758,FLM:386975}. 
In general, if an object or a process can be described with a single fractal dimension, it is called a monofractal, otherwise it is labelled as a multifractal, and an infinite number of dimensions could be employed to describe such processes \cite{pesin2008dimension,engelking,Hentschel1983435}. 
Systems that show different local distributions for its elements or some of its properties can be commonly found  in nature, such as the growth probability distribution of a diffusion-limited aggregation process (DLA) \cite{witten1981diffusion}, the  energy dissipation distribution in a network of resistors \cite{PhysRevE.61.R3283}, the variability in human behaviour \cite{Ihlen2013633}, and the soil's particle-size distribution \cite{Miranda2006373}.

In the case of cities it had been extensively argued \cite{chen2013multifractal,encarnaccao2012fractal,appleby,ArizaVillaverde20131,Hu2012161} that a single fractal dimension is not enough to describe its complex nature. 
In this paper we perform the multifractal analysis of the historical evolution of the London street network, by looking at nine digitised maps ranging from 1786 to 2010.

We found that as the city grows, the street network progressively fills the available space contained within the green belt, and thus creating over time a more homogeneous pattern.
This process results in the gradual loss of multifractality, in the sense that in the most recent time layers, the London street network  could be described by a single fractal exponent.

We argue that this multifractal to monofractal transition could be the result of the imposition of the statutory green belt around London in 1935 (1938 Act) \cite{guidance19952}, leading to a condensation phenomena in which the city fills the remaining space delimited by the boundaries. 
We test our conclusions by introducing and analysing the growth of a diffusion limited aggregation model (DLA hereafter), constrained by a non-permeable circular barrier.

\subsection{\label{ssec:level1} Dataset}

\texttt{\begin{figure*}[ht]
  \includegraphics[width=0.6\linewidth]{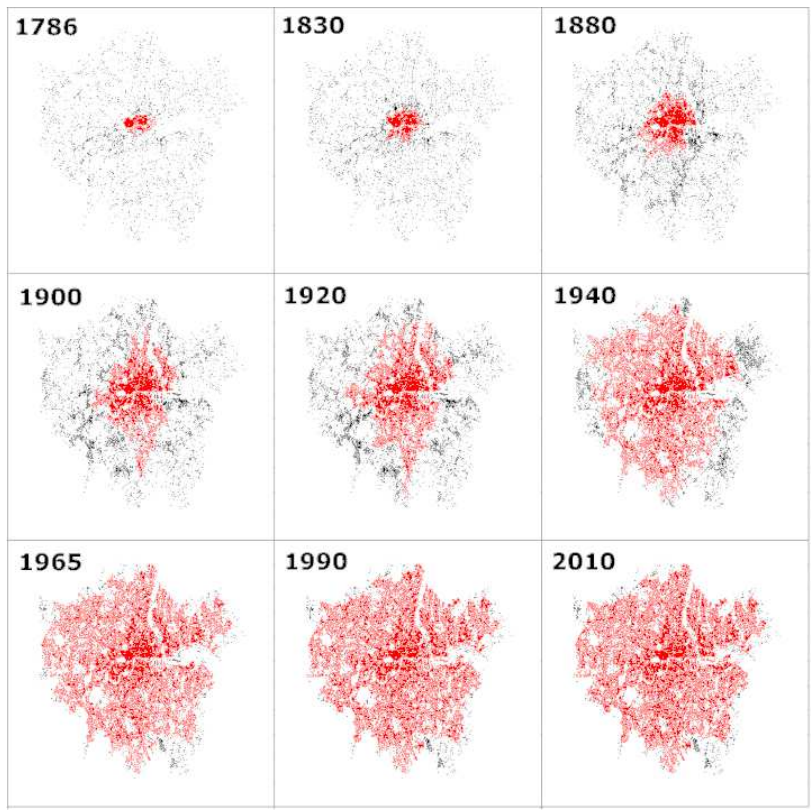}
  \caption{(Color online) GLA street intersection maps over time. In red we highlight the street intersection maps (SIPPs in the text) for the urbanised area as defined by the methodology employed in the text. }
  \label{Is11}
\end{figure*}}

In this work we employ a unique dataset, consisting of the digitised version of nine historical maps for the area contained in the Greater London Authority (GLA), from 1786 until 2010.
This dataset has already been studied in several works where further details can be found \cite{mas2013,masuccispatial,PhysRevE.89.012805}.

From each of the historical GLA maps, we extract the street network corresponding to only the urbanised area as defined by the methodology described in \cite{masarx}.
We then derive the street intersection map for each of the nine points in time, see Fig.~\ref{Is11}.
Our analysis is then reduced to the study of the multifractal properties of street intersection point patterns (SIPP) embedded in the two-dimensional Euclidean space.

Though multifractal analysis of point patterns is commonly used in many fields (cell colonies, electrochemical depositions, galaxy clusters distribution), to our knowledge this approach has never been used before in urban studies.
We argue that for the statistical analysis, it is equivalent to use the street intersections as it is to use the street segments. 
This is the case because the degree distribution of an urban street network is nearly Poissonian, and hence the average degree for the street connectivity and its variance are well defined quantities. 

Moreover, the SIPP has been proved to be a  quite resilient structure over time. 
Much of the original structure of the so-called the City of London at 1786 is still at place at present times. 
This situation can be observed for almost any major city in the world, which makes this approach a very good one to perform a historical analysis of cities.  

\section{\label{ssec:level2} Multifractal Measures}

\subsection{\label{sssec:level1}Review}
In this section we provide a brief review of multifractal measure theory. 
The uninterested reader could skip this part and go straight to Sec.\ref{ssec:level3}, where we summarise its applications to street networks.
A brief review of basic fractal theory is given in the Appendix \textit{A}.

Let us recall that when a monofractal object is characterised through a single global exponent, this is computed by looking at the distribution density of a specific measure, and it is implicitly assumed that this distribution is uniform \cite{meakin1998fractals}. 
Nevertheless, in real systems there exist many other intensive quantities of interest, such as the temperature, the hardness of a material, the electric field, and the density itself among others, whose distribution is non-uniform, in addition to sometimes presenting discontinuities at different scales.

This led to the definition of a local fractal dimension for each of the different regions or subsets of the system \cite{PhysRevA.32.2364}.
 
In growing systems, these heterogeneities are observed in the different growth probabilities for different regions.
For example, the DLA has different growth probabilities at the tips than at the \emph{fjords}. 
These sort of systems have been extensively simulated \cite{stanley1988multifractal,mandelbrot1990potential} through hit probabilities. 
A probability distribution function can hence be constructed from the probabilities $p_i$ of growth at location $i$, or from any other normalised measure that sums to one in the whole structure.
The multifractality of the object can hence be described through the distribution of the above-mentioned intensive measure.

Let us denote by $\mu(\boldsymbol{x})$, the value of a measure $\mu$ at position $\boldsymbol{x}$.
The amount of this measure within a volume $V(\epsilon,\boldsymbol{x})$ in the vicinity of $\boldsymbol{x}$ given by $\epsilon$, is defined by \cite{meakin1998fractals}
 \begin{equation} 
 \mu_\epsilon(\boldsymbol{x})=\int_{V(\epsilon,\boldsymbol{x})}{\mu(\boldsymbol{y})d\boldsymbol{y}}.
 \label{eq1}
 \end{equation}

Note that on a typical monofractal structure, this measure is homogeneous, i.e., 
\begin{equation} 
 \mu(\epsilon)\sim\epsilon^{-D}.
 \label{eq2}
 \end{equation} 
 where $D$ is the fractal dimension.

For multifractal objects the measure $\mu$ is different at different locations, hence the space can be subdivided into regions around neighbourhoods $\boldsymbol{x_i}$, where a distribution for each region can be obtained. If the measure is normalised, such a distribution can be constructed using $P_i(\epsilon)$ as follows

\be
P_i(\epsilon)= \mu'_\epsilon(\boldsymbol{x_i})= \frac{\mu_\epsilon(\boldsymbol{x_i})}{\int{\mu(\boldsymbol{y})d\boldsymbol{y}}}.
\label{eq3}
\ee

Given the inhomogeneity of the monofractal, and the infinitely many singularities of the distribution, for a multifractal Eq.~(\ref{eq2}) becomes \cite{PhysRevA.33.1141,Tél1989155}
\begin{equation}
P_i(\epsilon)\sim\epsilon^{\alpha_i}.
\label{eq4}
\end{equation} 
where $\alpha_i$ corresponds to the strength of the local singularity, and it is referred to as the \textit{Lipshitz-H\"{o}lder exponent} \cite{119727}. 

This exponent is not unique, as many other boxes or subdivisions of the space could have the same $\alpha_i$. 
The number of boxes with the same $\alpha_i$ value is given by \cite{PhysRevA.33.1141}
\begin{equation}
\mu(\alpha_i,\epsilon)=\epsilon^{-f(\alpha_i)}.
\label{eq5}
\end{equation} 
where the function $f(\alpha_i)$ is the fractal dimension of the set of subdivisions with singularity strength $\alpha_i$.

\begin{figure}[ht]
  \includegraphics[width=0.5\linewidth]{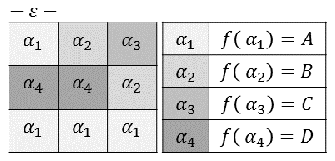}
  \caption{(Color online) Schematic representation for the distribution of the $\alpha - f(\alpha)$ pair. Each set of $\alpha_i$ has its own dimension. In this hypothetical example, the dimension A should be the highest, as it covers more space (four tiles) than the others. In contrast, the dimension C is zero, as it corresponds to a single point.}\label{alphaexample}
\end{figure}

To characterise a multifractal system, we therefore need to fully specify the  measures $\alpha_i$ and $f(\alpha_i)$. In Fig.~\ref{alphaexample}, we give a simple representation on how the different singularity measures $\alpha_i$ of the boxes of size $\epsilon$ are related to the fractal dimensions $f(\alpha_i)$. Note that if the box has a unique measure, the fractal dimension for that box will be zero.

Given that we have a distribution function $P_i(\epsilon)$ for the intensity measures in the $\epsilon$-region of a system, we can characterise this function via its moments \cite{PhysRevA.33.1141}
\begin{equation}
Z_q(\epsilon)=\sum_iP_i(\epsilon)^q.
\label{Zq}
\end{equation}

If $q=0$, we obtain the number of boxes $N(\epsilon)$ of size $\epsilon$ needed to cover the system
$Z_0(\epsilon)=N(\epsilon)\sim \epsilon^{-D}$.
In addition, via the exponent $q$ we can select different regions of the distribution $P_i$. If $q >> 0$ the regions with the largest values of $P_i$ will dominate the measure; similarly, if $q << 0$ the regions with the lowest values of $P_i$ will be the ones that will dominate the measure. 
Given that $\alpha_i$ is the scaling exponent of $P_i(\epsilon)$ with respect to $\epsilon$, $Z_q(\epsilon)$ in Eq.~(\ref{Zq}) should also obey a scaling relationship of the type
\begin{equation}
Z_q(\epsilon)=\sum_iP_i(\epsilon)^q\sim(\epsilon)^{-\tau(q)}.
\label{eq7}
\end{equation}
where ${\tau(q)}$ corresponds to the \textit{mass exponent} and can be defined as
\begin{equation}
\tau(q)=q\alpha(q)-f(\alpha(q)).
\label{tau_q}
\end{equation}
Eq.~(\ref{tau_q}) is usually written in terms of the \textit{generalised fractal dimensions} $D_q$ introduced by \cite{Hentschel1983435}  as
\begin{equation}
\tau(q)=(q-1)D_q.
\label{tau_qDq}
\end{equation}
where
\begin{equation}
D_q=\frac{1}{q-1}\lim_{\epsilon\rightarrow 0} \frac{\log(Z_q(\epsilon))}{\log(\epsilon)}.
\label{eqDq_lim}
\end{equation}
Eq.~(\ref{eqDq_lim}) allows us to recover three well-known dimensions \cite{grassberger1985generalizations}: for $q=0, D_0 $ corresponds to the capacity (box-counting) dimension; for $q=1, D_1 $ corresponds to the information (Shannon) dimension; and for $q=2, D_2 $ corresponds to the correlation dimension.  

If $D_q$ is known, the values for $\alpha$ and for $f(\alpha)$ can be obtained from Eq.~(\ref{tau_q}), (\ref{tau_qDq}) and (\ref{eqDq_lim})  as follows
\begin{equation}
\label{alpha1}
\alpha(q) = \frac{d}{dq}((q-1)D_q),
\end{equation}
 
\begin{equation}
q = \frac{d}{d\alpha}(f(\alpha(q))).
\label{q_d}
\end{equation}
Eq.~(\ref{q_d}) indicates that the maximum of $f(\alpha)$ is given by $f(\alpha(0))$, while all the other regions in the system have smaller dimensions. The definition of $D_q$ implies $D_q<D$ where $D$ is the fractal dimension of the substrate where our system is embedded, or simply the Euclidean dimension in this case.

\subsection{\label{ssec:level3}Applications to road networks}

Each one of the multifractal measures has a spatial meaning in the context of road networks.
That is of particular relevance in our case, as it allows us to make quantitative comparisons between the different SIPP studied. 

Multifractals systems are characterised by a decreasing function of the generalised dimension $D_q$ (Eq.~(\ref{eqDq_lim})), while for monofractals $D_q$ behaves as a constant function, and the distribution $f(\alpha_i)$ and the measures $\alpha_i$ are also constants..

First, let us focus on the three particular dimensions $D_q$, corresponding to $q\in \{0,1,2\}$.
Appendix \textit{B} outlines the methodology step by step to compute these measures.

\subsubsection{\label{sssec:level31}Capacity (box-counting) dimension: $D_0$}

This has been the preferred quantity identified as the fractal dimension, and it gives the probability of finding an intersection in an $\epsilon$-box, hence giving a sense of how the city occupies the space.

A high value for $D_0$, for example  1.9, indicates that almost all the boxes covering the city contain at least one intersection. 
Nevertheless, this measure is completely independent of the total mass found in each $\epsilon$-box, which means that this dimension tells us nothing about the intersection density distribution of the SIPP. 
For example, two different SIPP could have approximatively the same $D_0$ while their physical layout might be completely different.   
 
\subsubsection{\label{sssec:level32}Information dimension: $D_1$}
For $q=1$, we calculate Eq.~(\ref{eqDq_lim}) when $q\rightarrow 1$, and obtain the following equation:

\begin{equation}
D_1=\lim_{\epsilon\rightarrow 0} \frac{-\Sigma_i P_i log P_i}{-\log(\epsilon)}.
\label{eq19}
\end{equation}
where the numerator is Shannon's entropy, which measures  the amount of information related to the density distribution of the SIPP.
Eq.~(\ref{eq19}) is an expression for the unevenness of the point density distribution in the $\epsilon$-boxes \cite{grassberger1985generalizations}.
Higher values of $D_1$  reflect a more uniform intersection distribution over the whole city, which in turn, could represent either a saturation of the space or the fact that the city has a regular structure in terms of SIPP.

\subsubsection{\label{sssec:level33}Correlation dimension: $D_2$}

When $q=2$, we obtain the correlation dimension $D_2$.
This is a dimension of the correlations between pairs of intersections, $ I1 $ and $ I2 $, i.e., it is the probability that $ I1 $ and $ I2 $ lie within the same $\epsilon$-box.

\begin{equation}
D_2=\lim_{\epsilon\rightarrow 0} \frac{\log\Sigma_i P_i^2}{\log(\epsilon)}.
\label{eqcorrelation}
\end{equation}
 
In this way the correlation dimension  gives more accurate results than the box-counting dimension for spatial objects, as it explicitly takes into account the spatial structure of the SIPP. 
A low value for $D_2$   reflects a spare intersection structure, while large values for the same quantity represent more compact and ordered structures.

\subsubsection{\label{sssec:level34}The generalised dimension $D_q$ }

For the remnant  multifractal dimensions related to other $q$ values, we have to keep in mind that they are defined via  $P_i(\epsilon)$ (Eq.~(\ref{eq3})), which is the probability of finding an intersection in the $i_{th} \epsilon$-box.

This differs to measuring the multifractal spectra for the DLA \cite{vicsek1990multifractal}, which considers the growth probability on the $\epsilon$-box.
However, the probability of finding an intersection could be related to the growth probability, given that the areas in the city with larger $P_i$ have a lower growth probability, and vice versa. 
As stated in Sec. \ref{sssec:level1}, $q$ allows us to select different regions according to their relative $P_i$ values. 
In this sense, it can be considered a resolution parameter.

In terms of the SIPP, positive $q$ values correspond to multifractal measures for areas with more intersections, while negative ones correspond to measures for areas with less intersections.

\subsubsection{\label{sssec:level35}The mass exponent $\tau$}

As stated above, for different values of $q$, we obtain different probabilities of finding an intersection. 
The mass exponent $\tau(q)$ reflects this through Eq.~(\ref{eq7}). 

A SIPP with a large $\tau(q)$ will have more intersections at the given resolution $q$ than a SIPP with a smaller $\tau(q)$.
 
\subsubsection{\label{sssec:level36}The singularity exponent $\alpha(q)$}

For different values of $q$, each region of a SIPP displays a different probability $P_i$ of finding an intersection on it. 
As explained in Sec. \ref{ssec:level2}, it is possible that different regions share the same $P_i$, or at least very similar ones. 
In such a case, the singularity exponent $\alpha(q)$ is a measure that  captures the variety of intersection densities in a SIPP at resolution $q$. 
A high value for $\alpha(q)$ means that a high variety of densities at that particular resolution is found.

\subsubsection{\label{sssec:level37}The multifractal spectrum $f(\alpha(q))$}
The multifractal spectrum $f(\alpha(q))$ represents the dimension $f$ of the set of regions which display similar $\alpha(q)$ values.
 While the $D_q$ represents the dimension obtained by examining the distribution of the intersections over the whole SIPP, the $f(\alpha(q))$ is the dimension obtained over the different regions which display the same $\alpha$. 
 The plot $f(\alpha(q))$ \textit{vs} $\alpha(q)$ gives us a global picture of the whole structure of the SIPP in terms of the extension and variety of the different intersection distributions.

\section{\label{sec:level5}Results}
\subsection{London street network}
In order to show that London's street network  evolves from a multifractal SIPP, to a structure whose multifractal characteristics are almost lost, we  plot and analyse the corresponding multifractal measures.
To do so, we apply equations (\ref{Zq}) to (\ref{eqDq_lim}) to each one of the nine SIPP obtained from the historical data set,  having considered the constraints on the range for the $q$ parameter mentioned in Appendix \textit{B}. 

 \begin{table}[h] 
  \caption{ \label{s5t1}
  Main parameters obtained from the multifractal analysis of each SIPP. The range for the $q$ is not the same for each year due to restrictions outlined in the Appendix B}
 \begin{ruledtabular}
 \begin{tabular}{cccccc}
  $Year$ & $q$ & $D_0$ & $D_1$ & $D_2$ & $\alpha(0)$ \\ 
 \hline
 1786 & [-5.00,14.25] & 1.7959 & 1.7697 & 1.7467 & 1.8287 \\ 
 1830 & [-7.00,13.25] & 1.7926 & 1.7698 & 1.7448 & 1.8187 \\
 1880 & [-5.75,14.50] & 1.8196 & 1.7950 & 1.7722 & 1.8487 \\ 
 1900 & [-4.75,19.00] & 1.8434 & 1.8252 & 1.8109 & 1.8658 \\ 
 1920 & [-4.25,17.75] & 1.8602 & 1.8431 & 1.8322 & 1.8824 \\ 
 1940 & [-5.25,18.75] & 1.8699 & 1.8619 & 1.8586 & 1.8820 \\ 
 1965 & [-5.25,16.00] & 1.8850 & 1.8803 & 1.8780 & 1.8927 \\
 1990 & [-4.50,15.00] & 1.8851 & 1.8793 & 1.8766 & 1.8948 \\
 2010 & [-4.50,16.00] & 1.8913 & 1.8858 & 1.8842 & 1.9004 \\
 \end{tabular}
 \end{ruledtabular}
 \end{table}

The actual values for $q$ that have been taken into account, along with the main dimensions and $\alpha(0)$,  are shown in Tab. \ref{s5t1}. 
We can observe a consistent asymmetry between the $q^+$ and $q^-$ values, where the range for the positive values is much larger than the one encountered for the negative values. 
This means that for each SIPP, the number of significant areas with low probabilities of having an intersection in them, which are the ones magnified by $q^-$ values, is always smaller than the ones with high probabilities.  
Such a difference is confirmed by  other measures, and particularly by the multifractal spectra $\alpha(q)-f(\alpha(q))$, as we  show below.

\begin{figure*}[htb]
  \centering
  \includegraphics[width=.90\linewidth]{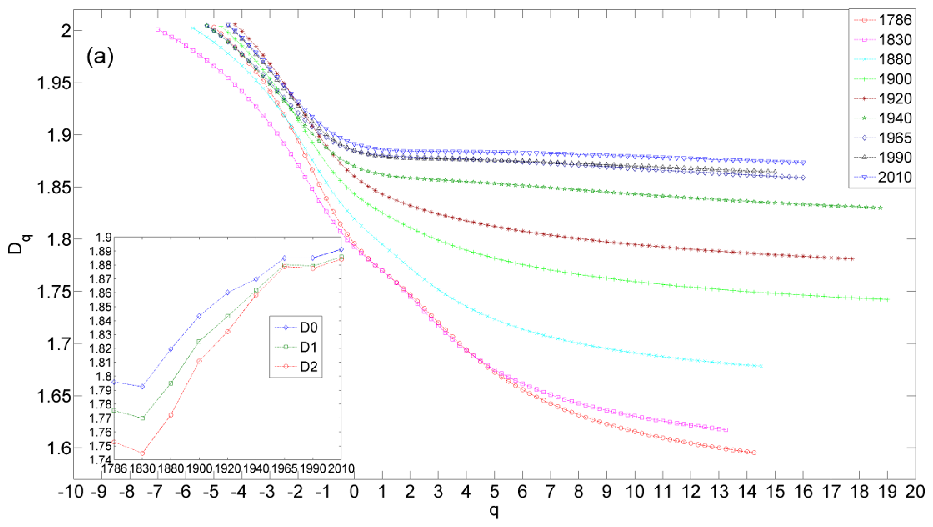}
  \includegraphics[width=.45\linewidth]{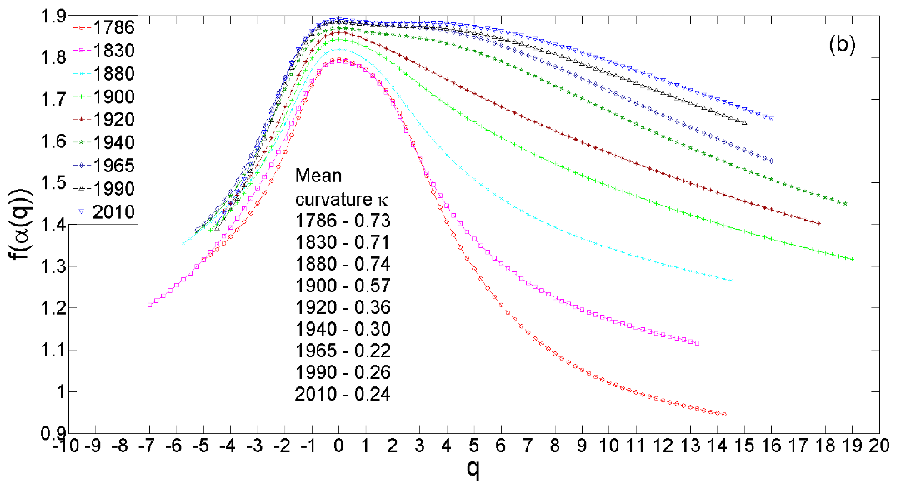}
  \includegraphics[width=.45\linewidth]{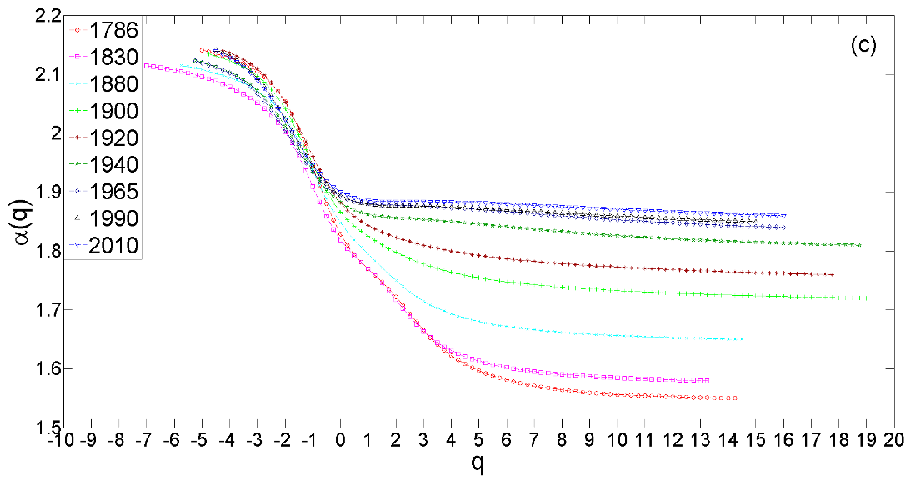}
  \includegraphics[width=.45\linewidth]{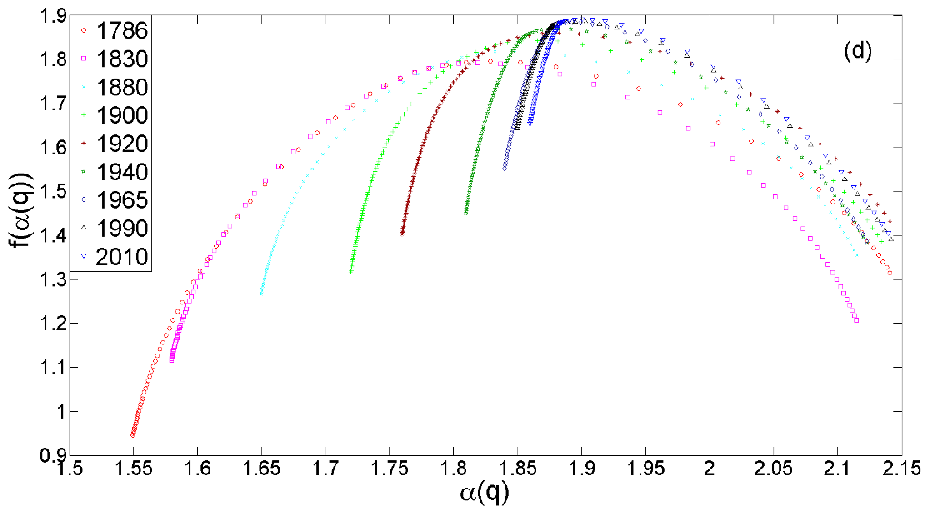}
  \includegraphics[width=.45\linewidth]{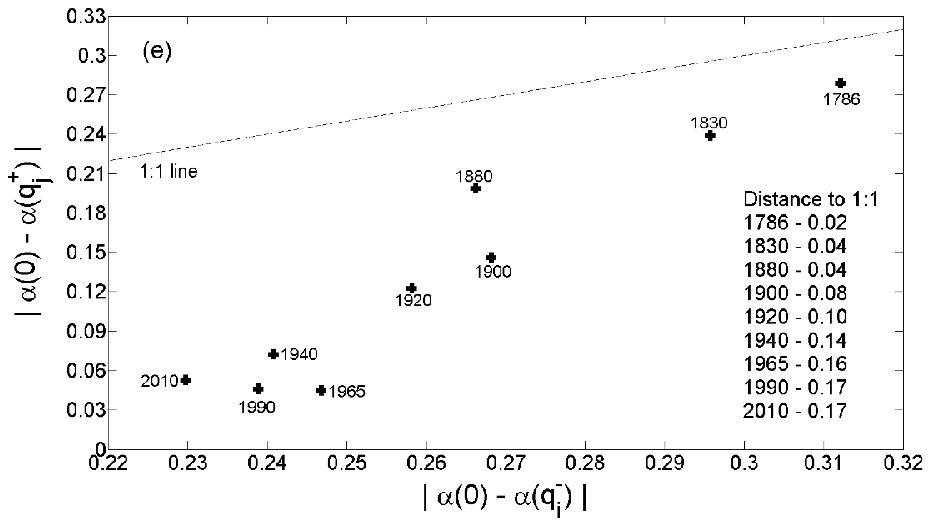}
  
  \caption{\textbf{(Color online) Multifractal Analysis} for all the years studied. The differences in value between these measures follow a pattern that imply that a single dimension is enough to characterise the SIPP, or, in other words, the structure is undergoing a transition from a multifractal to a monofractal structure.}\label{MFA}
\end{figure*}

In Fig.~\ref{MFA}, \textit{panel a}, we show $D_q$ as a function of $q$.
We can see how for the oldest SIPPs the $D_q$ values for high $q$ are quite low. 
This relates to the fact that  the old London street network is scarce with dense areas.
The increasing of $D_q$ for high $q$ during the years shows us how these dense areas became more important during the SIPP evolution.
The evolution of these areas can be tracked looking at the distance of $D_q$ between $q^+$ values from one year to another. 
As the gap between two lines is larger, the morphological difference between structures is also larger. 
As an example, for $q=5$, the SIPP distribution between 1880 and 1900 is less similar than the one between 1900 and 1920; the differences for all $q$ values between the last three years is almost zero, i.e., it is not trivial to distinguish one from the other in terms of the distribution of its intersections.
This  is an evidence of how  the London's SIPP  evolved in time toward a condensation, as we better explain below. 
The inset plot at Fig.~\ref{MFA}, \textit{panel a}, shows the evolution of $D_0$, $D_1$ and $D_2$ during time. The clear difference between these values ($D_0>D_1>D_2$) from 1786 to 1920 is a clear signature of the system's multifractality during this period. From 1940, the nominal distance between $D_1$ and $D_2$ begins to decrease, until 2010 when is practically zero (0.001), while the distance from $D_0$ and $D_1$ at this same year, is only 0.005. This strongly suggests a decay in the SIPP multifractal characteristics over the years, almost to the point to be able to describe the last three SIPPs with a single generalised dimension,  \textit{ie}, $D_2$. 

Such a phenomenology is related to a homogenization process for the SIPPs, possibly related to a condensation phenomena, as we show below.

The mass exponent $\tau(q)$ as a function of $q$  has the expected monotonic growing behaviour, as the SIPP number of intersections increases year after year. 
Until $q=3$, $\tau(q)$ is practically the same for all years. As the $q$ values increase, the areas with more intersections are the ones enhanced, and it is there when the differences between $\tau(q)$ values  emerge. 
From Eq.~(\ref{tau_q}), we know that $\tau(0)=-D_q$, $\tau(1)=0$ and $\tau(2)=D_q$. 
Then, as it has been argued, different values for $D_q$ imply multifractality. 
As the non-linear behaviour for $\tau(q)$ would, in turn, imply multifractality, we test the linearity for the $\tau(q)$ for each year, fitting it against a linear function and looking at the norm of the residuals of such fit. 
If the  norm is equal to zero, then the function can be characterised as linear. 
For the first three years, we have higher norms (all $> 1$), but as we move in time, this norm is approaching  zero, a situation that implies lost of multifractaliy during the growth of the London SIPP. \\

In \textit{panel b} of Fig.~\ref{MFA}, we show the multifractal spectrum $f(\alpha(q))$ as a function of $q$, for the  SIPPs.
The function $f(\alpha(q))$ conveys information with respect to the distribution of the intersections, in the sense that it relates the different areas enhanced by $q$ with a dimensional quantity. 

We observe that for $q^-$, the curve follows the same behaviour for all the years, indicating that the number of areas with low probability of having an intersection remains more or less constant during the years.

On the other hand, we observe significant structural differences in the SIPP between the different years for $q^+$.

In particular, we observe a change in the concavity of the curve once we reach 1920, and for the most recent 3 points in time, the curves almost overlap at $q\approx5$.\\

Let us denote by $\kappa$ the mean curvature of $f(\alpha(q))$ in a closed neighbourhood of this point. 
$f(\alpha(q))$ has a maximum at $q=0$, and for a monofractal structure $f(\alpha(q<0))=f(\alpha(0))=f(\alpha(q>0))$, hence $\kappa=0$.
For a multifractal structure, we expect to find $\kappa\neq 0$.
The plot tells us how the value of $\kappa$ decreases with the evolution of the network, reaching a stable low point for the past 50 years.

 In \textit{panel c} of Fig.~\ref{MFA}, we show the behaviour of the singularity exponent $\alpha(q)$ as a function of $q$.
 It represents the structure of the SIPPs in terms of the dominant probabilities at resolution $q$.
 Once again we observe that for $q^-$ the structure of the SIPP is very similar for all years, while for $q^+$ major differences between the years emerge. 
 For the earlier years, there are only a few areas with major intersection densities, and so the values for $\alpha(q)$ are low for $q^+$.
 On the other hand, for the latest years, this situation is inverted. 

 From $q=0$, the values for $\alpha(q)$ for the three most recent SIPPs are very similar, i.e. the distribution of probabilities is the same for all areas (less/more intersections), while for the rest of the SIPPs, the differences between these  values are greater, particularly for the first two ones. 

Nevertheless, from $q=7$, each of the values for the SIPPs tend to converge to a constant value, indicating that from this point the different spectra remain constant and hence the resolution limit for $q$ has been reached.

 In \textit{panel d} of Fig.~\ref{MFA}, we show the multifractal spectrum $f(\alpha(q))$ as a function of $\alpha(q)$.

As we already mentioned, each point of the plot shows the fractal dimension $f(\alpha(q))$ of the areas with the same singularity strength $\alpha$, at resolution $q$. 
 In this sense, it is important to note that $f(\alpha)$ does not represent the dimension of continuous regions, which tends to be a common misunderstanding. We observe that these curves have a maximum for different $\alpha(q)$ values. 
 These maximum values correspond to $D_0$, at $\alpha(0)$. 

It might then become clear that the $\alpha(q)$ values to the left of the maximum are the ones associated with $q>0$, while the ones to right are the ones associated with $q<0$. 
Keeping in mind that $D_q$ is a dimension for the whole structure, while $f(\alpha(q))$ is the dimension of a subset of such structures, the clear asymmetry of the curves reflects at some extent the asymmetry in the distribution between regions with large/small number of intersections. 
Then the differences between the spectra for areas with high density  of intersections are greater than the differences we find for the areas with small intersection densities.

The fact that the dimension $f(\alpha(q))$ is greater for the zones with $\alpha$ values associated with $q<0$, which are the areas with less number of intersections, is not a surprise. 
For example, for 1786 the $f(\alpha(q))$ dimension for the areas with larger number of intersections is very low; while the $f(\alpha(q))$ dimension for the areas with less intersections is larger. 

If the multifractal spectrum collapses to a single point, it means that the dimension $f(\alpha(q))$ becomes the same for all $\alpha(q)$, i.e. either all the regions in the structure share the same strength exponent or all the different strength exponents have the same dimension at all resolutions. 
In either case, this means that the structure is a monofractal. 

For the case of the SIPPs, it is clear that none of them collapse in such a way, nevertheless we can observe a transition from a wide curve in 1876 to a narrow one for the last two SIPPs. 
This indicates a progression from multifractal to monofractal characteristics. \\

The symmetry of the curve $\alpha(q)$ vs $f(\alpha(q))$  gives us information about the relationship between regions with different distributions. 
In \textit{panel e} of Fig.~\ref{MFA},  we show a symmetry measure for the multifractal spectrum. 
This is done by plotting the points $\{ |\alpha(0)-\alpha(q_i^-)|,|\alpha(0)-\alpha(q_j^+)|\}$ for each year, where $q_i$ and $q_j$ stands for the last/first valid $q$ value, and the 1:1 correspondence line. 
The distance from these points to this line is a visual tool to understand the asymmetry of the relative spectrum. 

This symmetric feature represents the balance between areas with more intersections and areas with less intersections. 

The actual distance for each point is shown in the inset of this same plate. It is clear that this distance increases as we move forward in time: for 1786 the point lies at a distance of 0.02, while for the last three years the distance $\sim 0.17$. 
This loss of symmetry, due to the increasing number of intersections, reflects the undergoing condensation phenomena in our study area, and once more, supports our hypothesis on the evolution of London from a multifractal to a monofractal. 
\\

\subsection{Model: Constrained DLA}

The multifractal measures show that London's street network grows while undergoing a process of morphological re-structuring that is manifested in fundamental differences in the distribution of the structures between different time periods. 
The behaviour of the network, with respect to the way it fills the space, follows an expected trend until 1920, and from this year an anomalous evolution seems to take place, with $D_1$ and $D_2$ starting to collapse to a single value. 
We argue that such a change comes as the result of two factors: the extensive diffusion of development from the core of the city to the periphery with massive suburbanisation and, after the mid-1950's, the effect of the actual implementation of a green belt around London, in order to contain urban sprawl.

In previous work we showed that the city undergoes a condensation process in its evolution as the result of the green belt \cite{mas2013}, here we show its effects on the inner structure of the city for the first time.

In order to test our hypothesis of the evolution of London from a multifractal to a monofractal structure due to the presence of the green belt, we introduce a multifractal diffusion model in a constrained plane, and we explore the effects of a restricted barrier on a growing multifractal structure. 
Our model is based on the classic DLA algorithm \cite{witten1981diffusion} modified in a way that mimics a green belt barrier set at a distance \textit{d} from a seed point at (0,0). 

\begin{figure*}[ht]
  \centering
  \includegraphics[width=.3\linewidth]{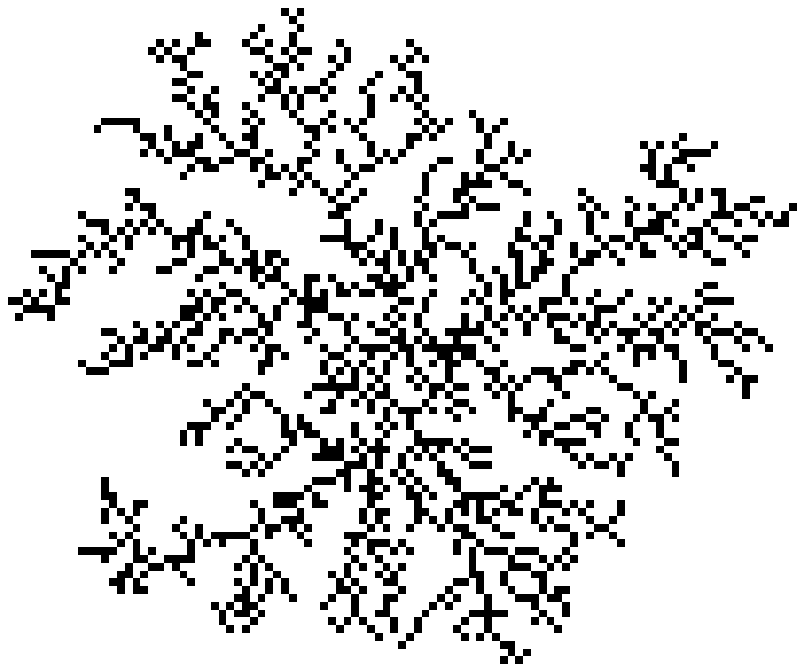}
    \includegraphics[width=.3\linewidth]{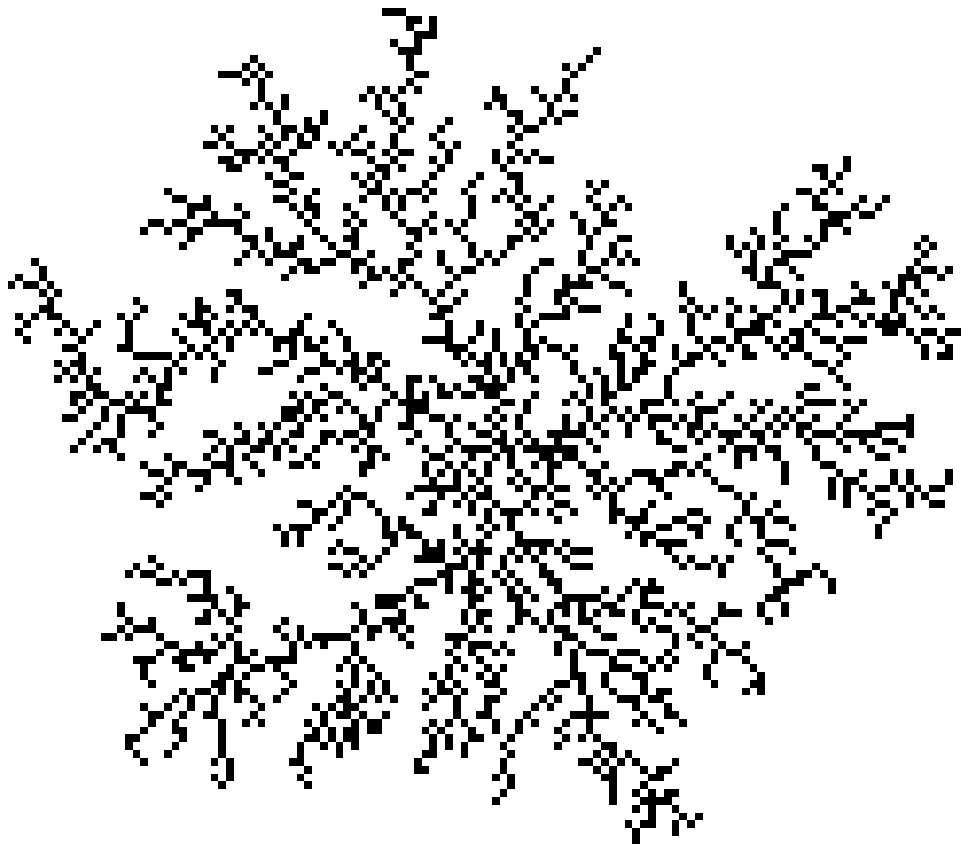}
    \includegraphics[width=.3\linewidth]{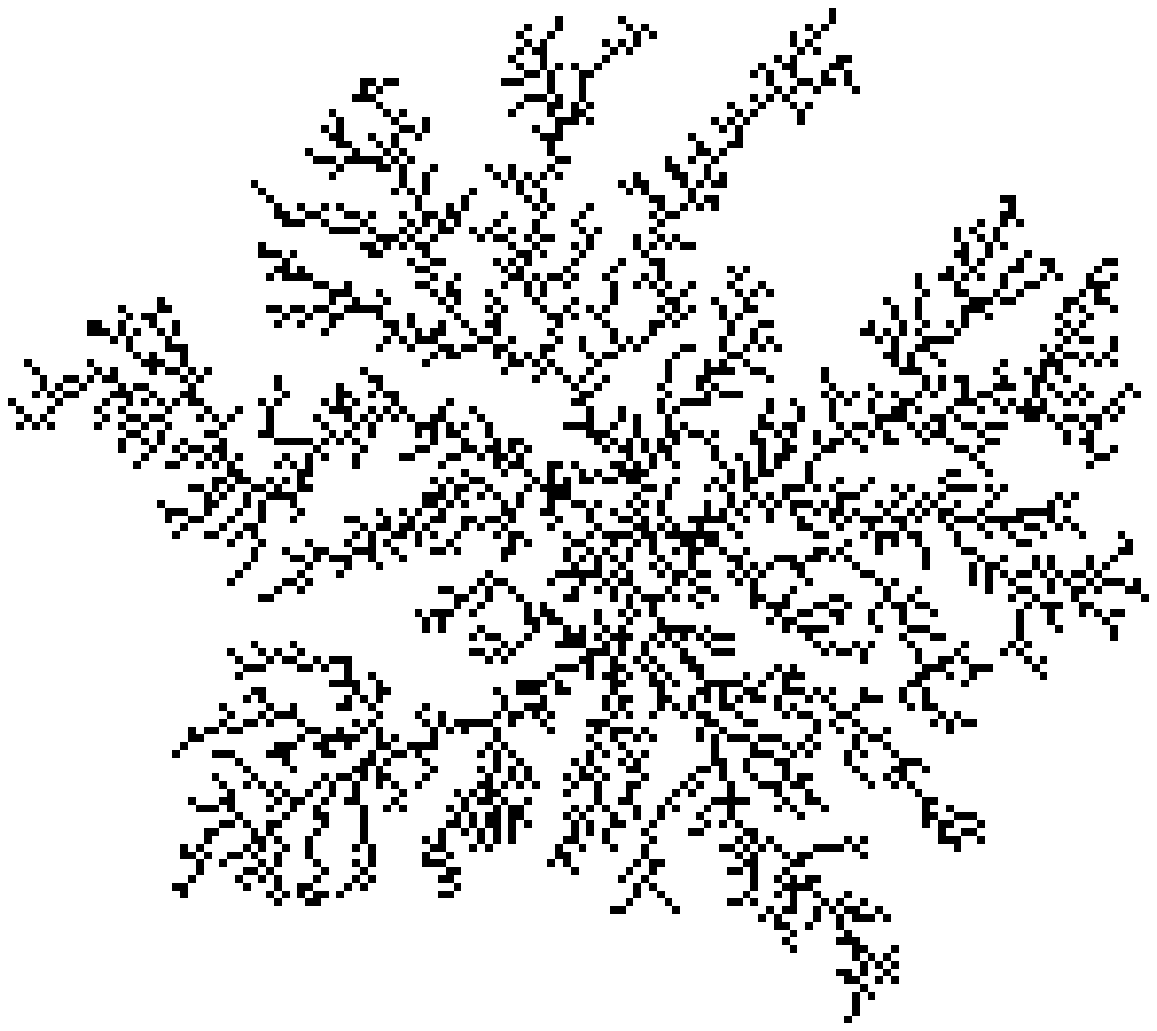}
    \includegraphics[width=.3\linewidth]{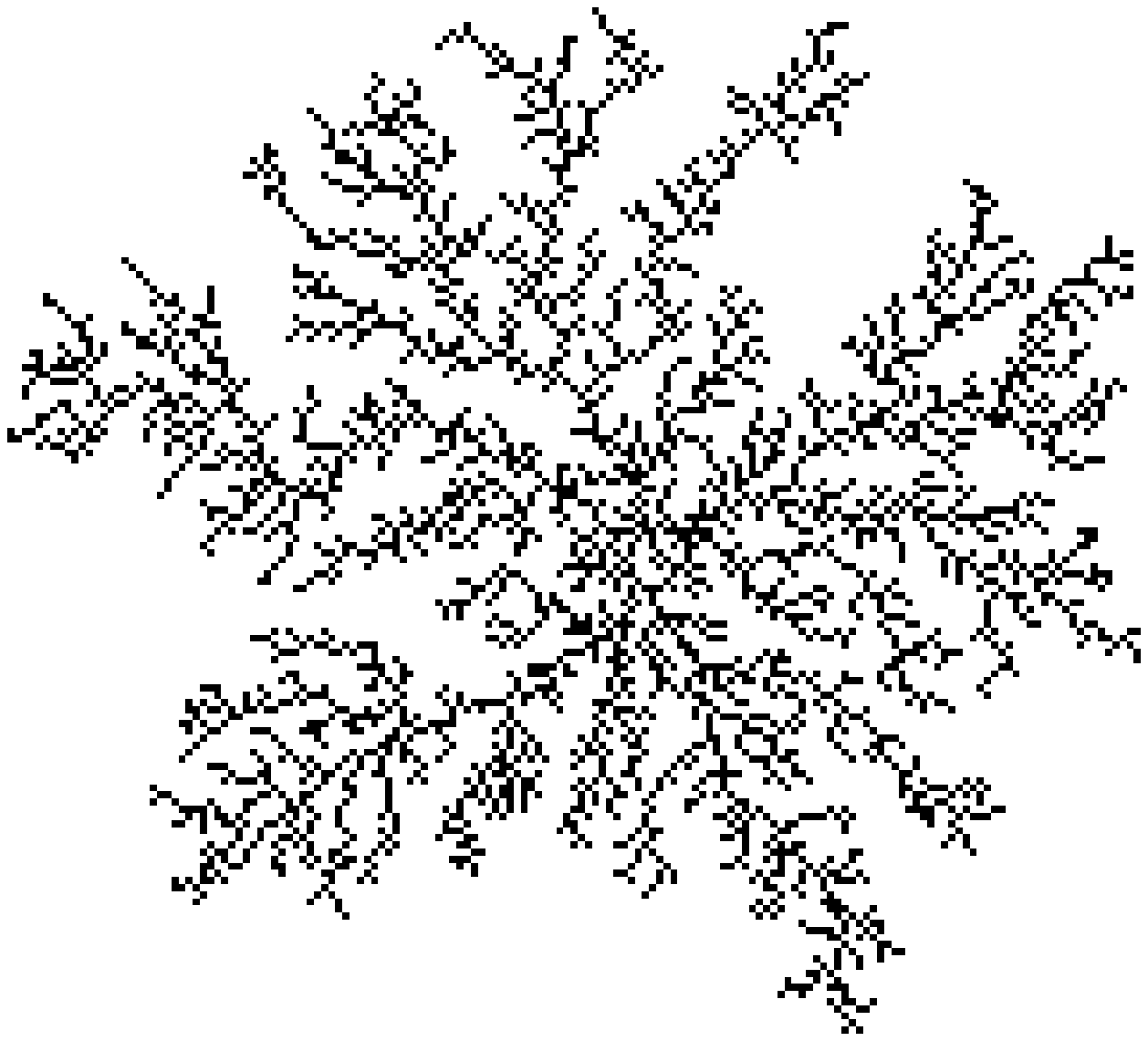}
    \includegraphics[width=.3\linewidth]{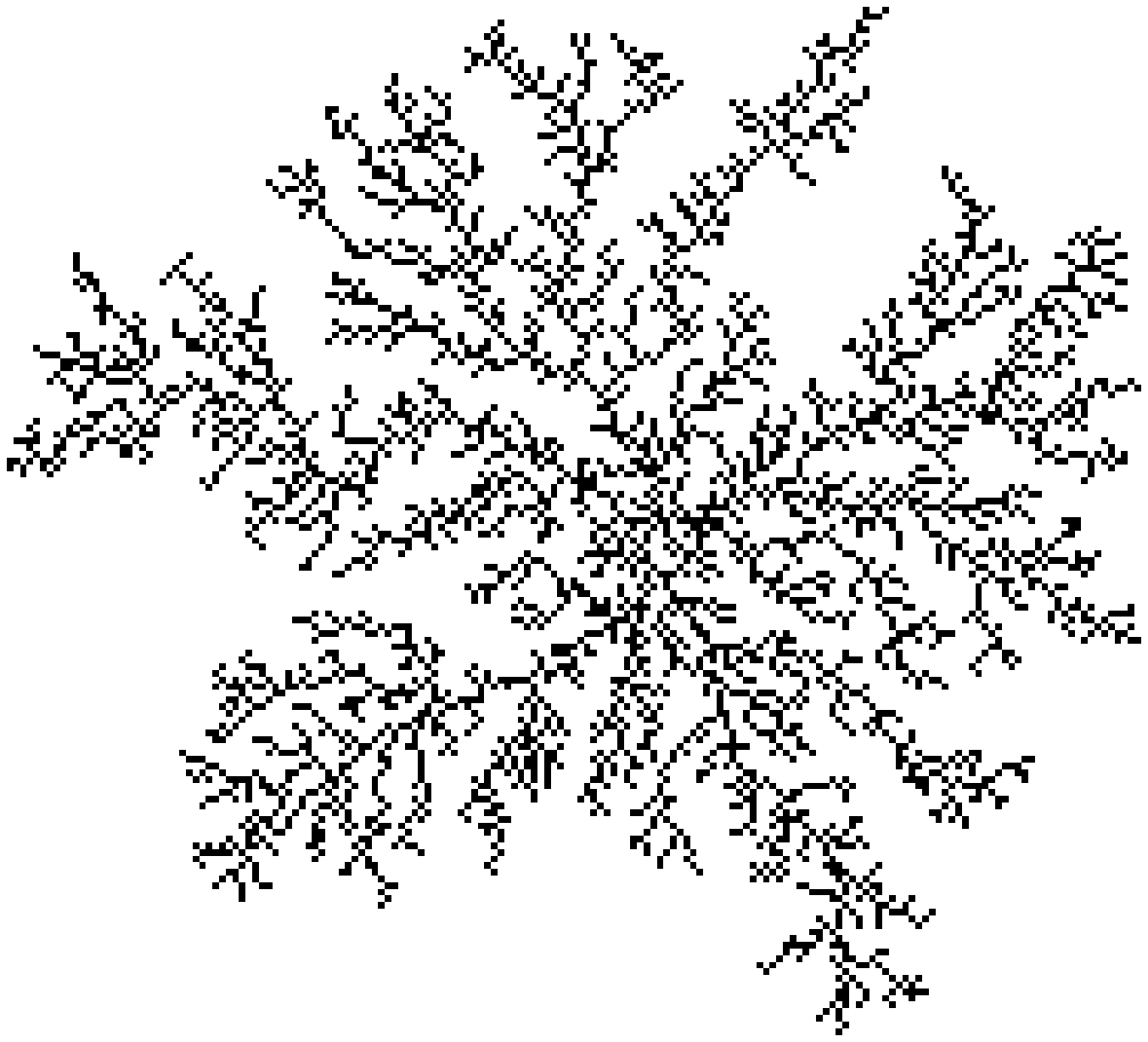}
    \includegraphics[width=.3\linewidth]{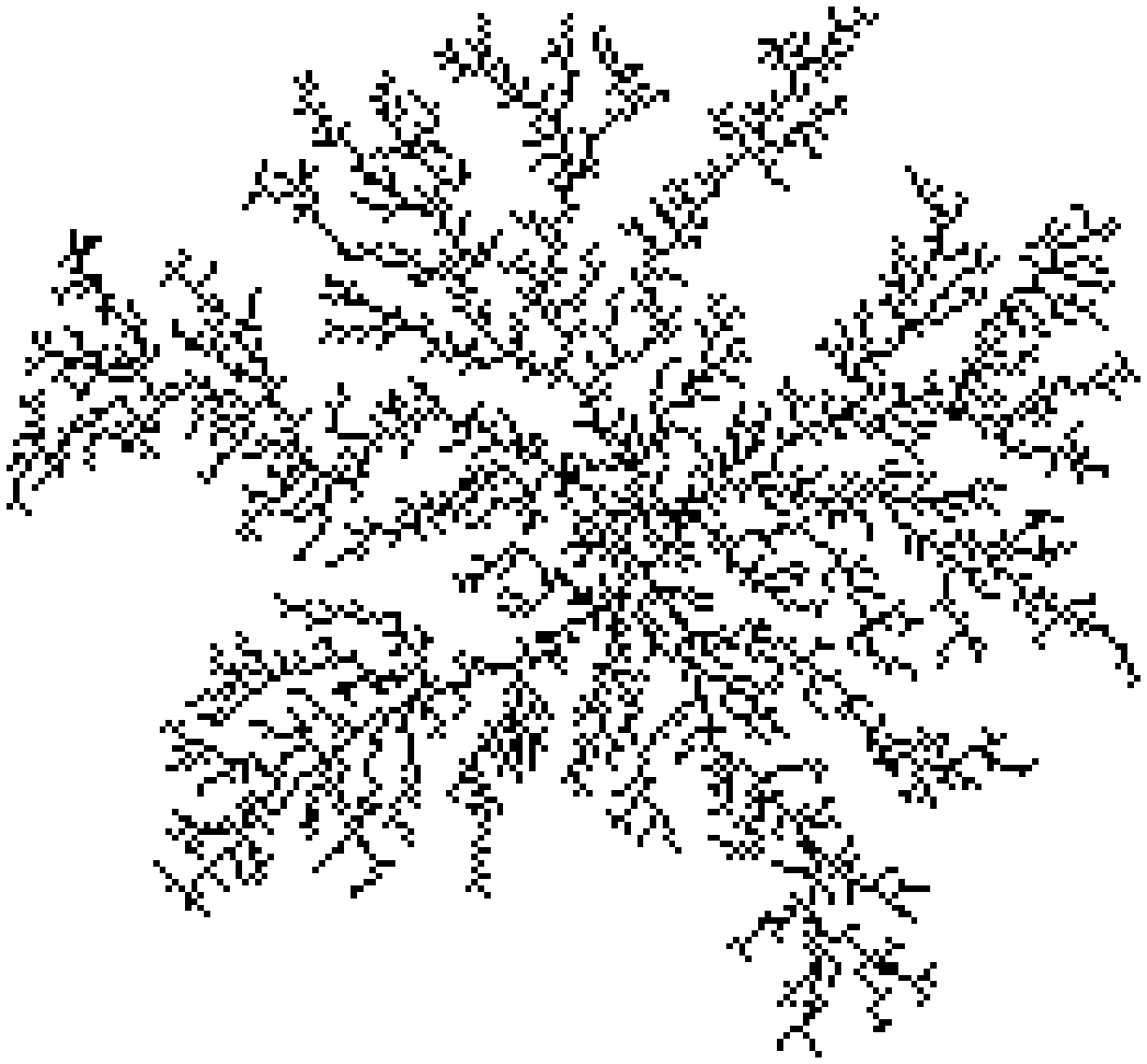}
     \includegraphics[width=.3\linewidth]{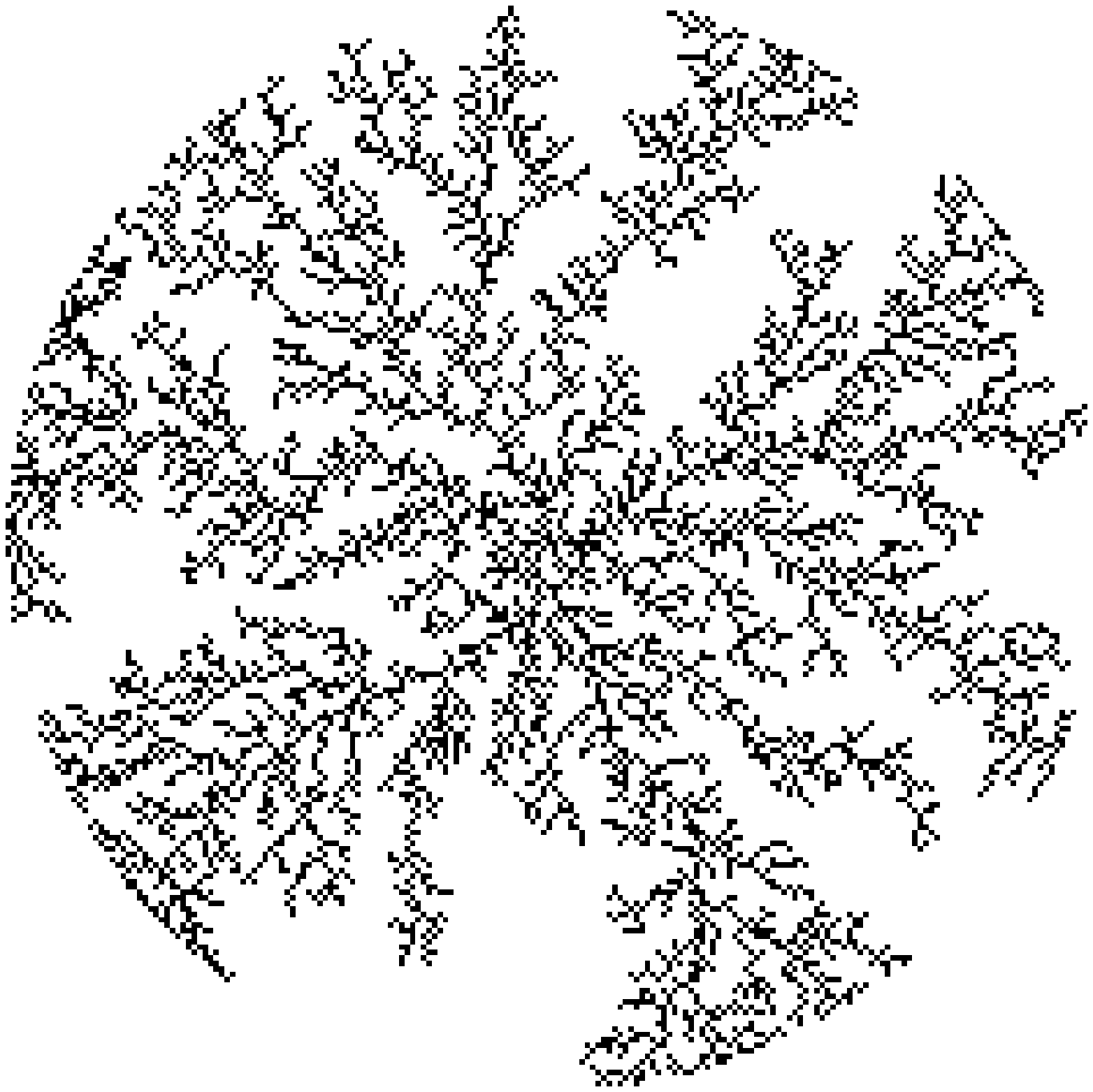}
    \includegraphics[width=.3\linewidth]{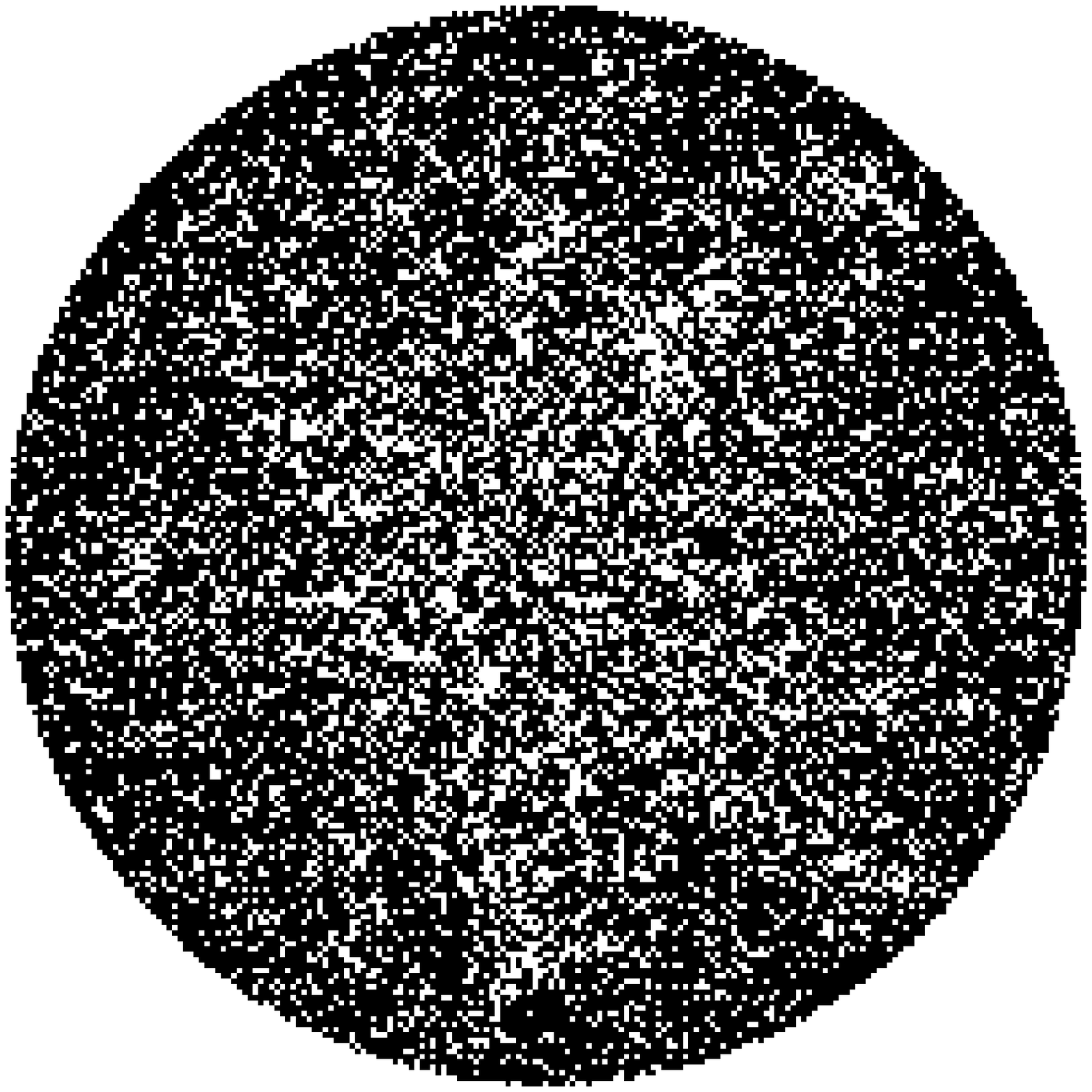}
    \includegraphics[width=.3\linewidth]{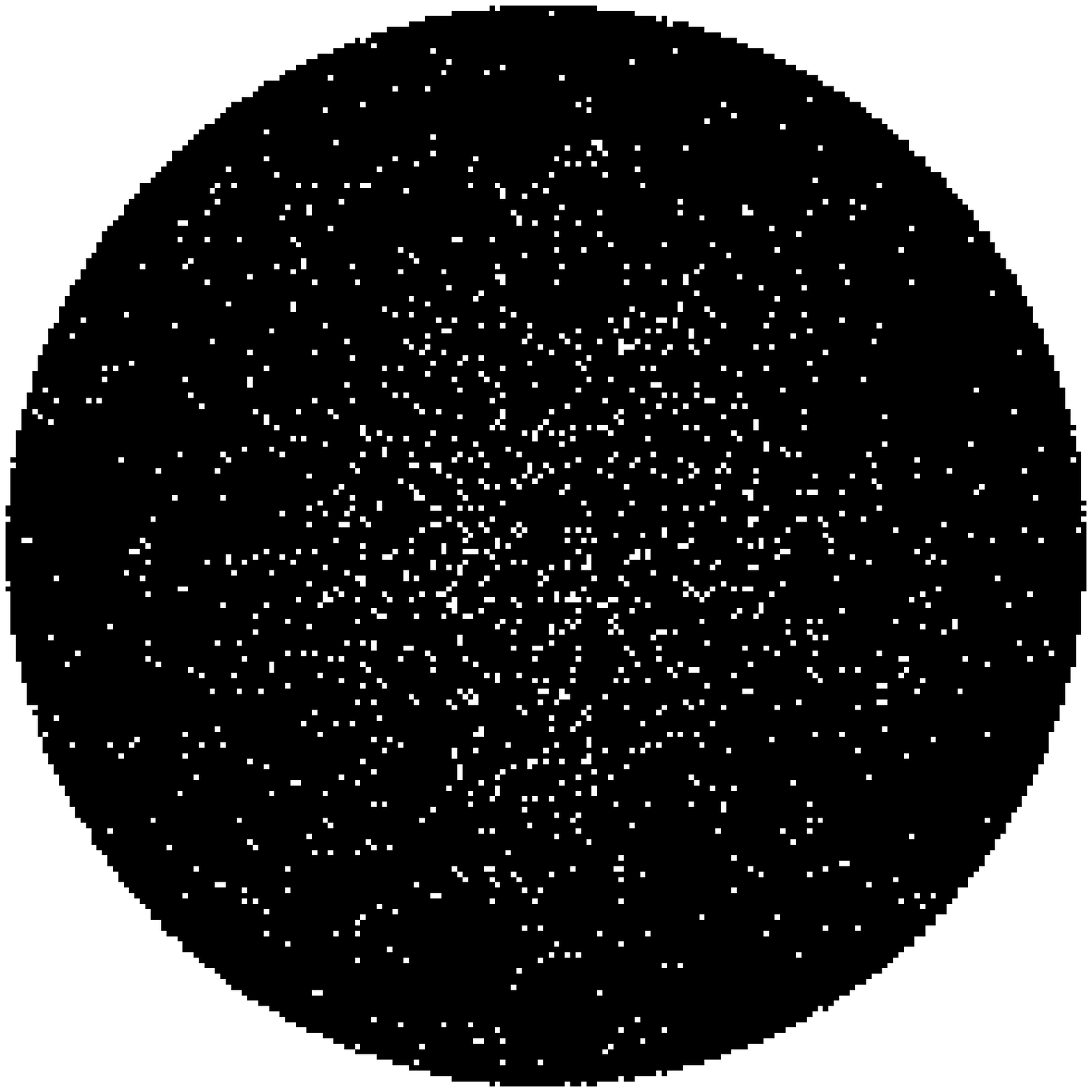}
  \caption{Evolution of the constrained DLA  model}
    \label{model2}
\end{figure*}

Given that the DLA structures are formed through a stochastic process, we perform 100 runs to derive some average measures, and since the historical data set used in this investigation consist of nine points in time, we construct a nine point time series $T$ for our model. 
Each point $T_i\in{T}$  is selected at one particular iteration, representing different stages of growth. \\
We define the barrier as a non-permeable circular perimeter $c$, with radius $r=100$ and centre at (0,0). 
The model follows exactly the same rules as in the DLA, until the wandering particles enter in contact with the green belt. 
At this stage, the particles are not allowed to cross the barrier and are forced to find a new position in the available space inside the barrier. 

The effect of the barrier on the DLA model can be observed from the sixth stage of growth onwards, see Fig.~\ref{model2}.

\begin{figure}[ht]

  \centering
      \includegraphics[width=.90\linewidth]{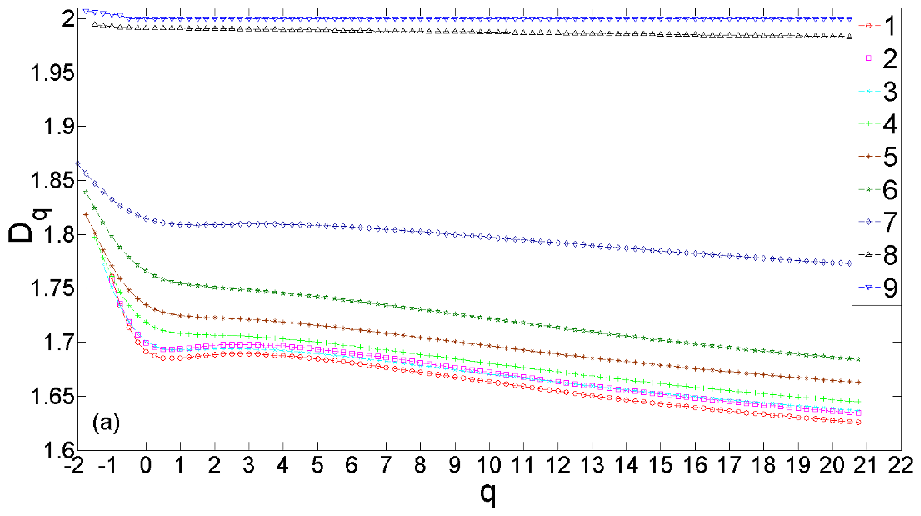}\\
      \includegraphics[width=.90\linewidth]{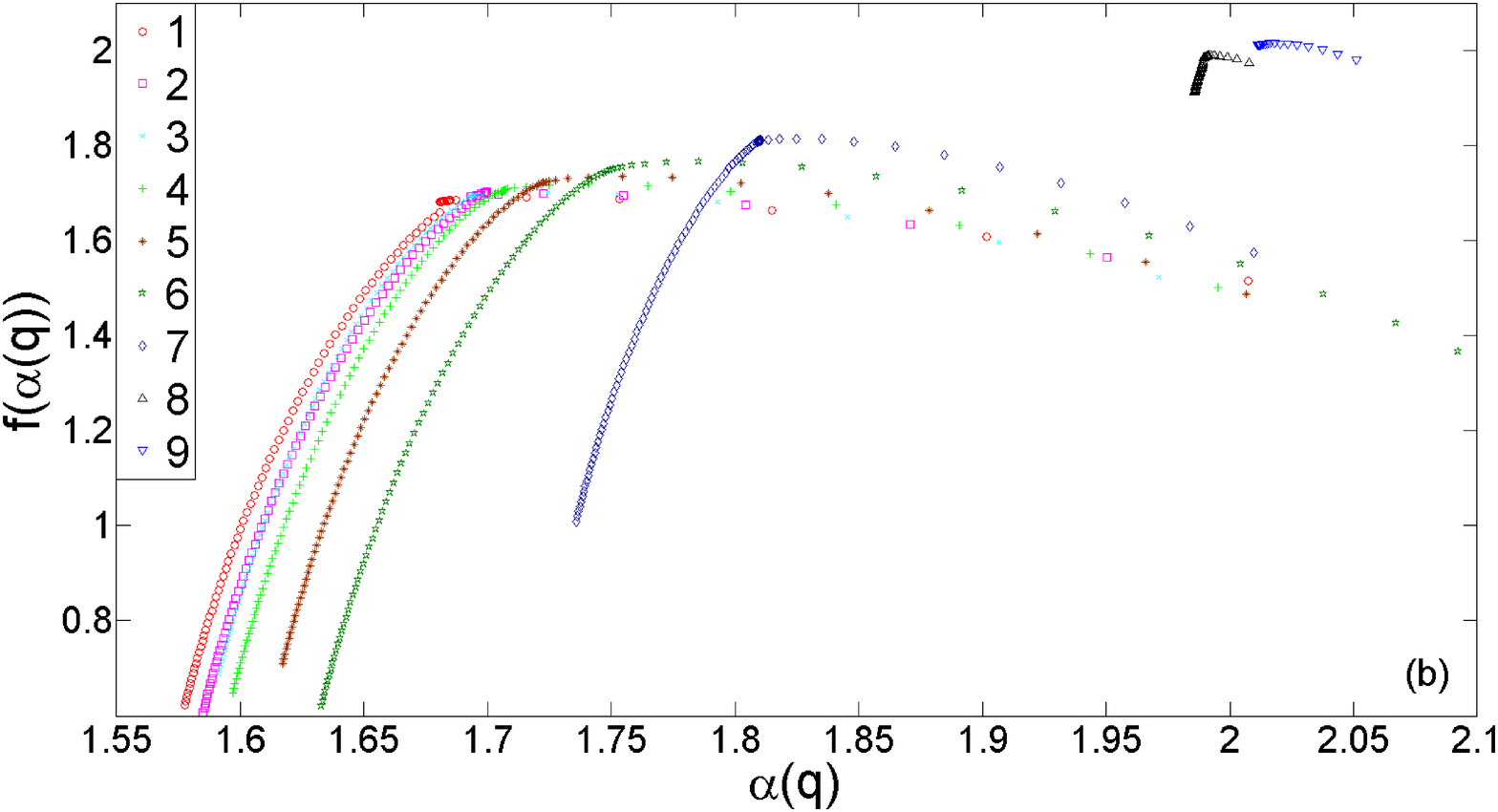}
   \caption{(Color online) a) $q-D_q$ plot. Until point 6, when the structure begins to interact with the green belt, all the points for both models depicts the same similar behaviour. From point 7 the effect of the barrier over the structure is clear. b) $\alpha(q)$ vs $f(\alpha(q))$ plot. The spectrum at points 8 and 9 are almost completely collapsed to a single point, which is an  indicator of a monofractal structure.}
   \label{Dqmodels}
\end{figure}

We then apply equations (\ref{Zq}) to (\ref{eqDq_lim}) to each aggregate in order to investigate the effects of the growth restriction on the model with respect to the inherent multifractal characteristics of the DLA (see Fig.~\ref{Dqmodels}). 
We observe that all the measures related to multifractality start to collapse toward monofractality after the sixth point.
In the last two stages in Fig.~\ref{model2}, the model is practically a filled 2-D circular plate, with a correlation dimension $D_2=2.0$.

\section{\label{sec:level6}Conclusions}

Several studies have already  showed that cities are multifractal objects, instead of fractals structures, as it was originally suggested. 
Here we  show that for London, this is only true up to a certain point in time, when the city starts a condensation process forcing the  new street intersections to be allocated in the available space inside a confined area, i.e., not in a sprawling fashion. 
For the three most recent maps examined in this analysis (1965, 1990, 2010) all our measures are very similar, reinforcing the idea  that a condensation process is in place and that it is reaching a stable phase. This implies that $\alpha_i$ in Eq.~(\ref{alpha1}) is equal $\forall$ \textit{q}, i.e. the probability of finding an intersection at any random area in the SIPP is the same.

We observed that such a multifractal to monofractal transition is not particularly smooth, in the sense that the differences in values get dramatically shortened from 1940 and 1965, which is the period when the green belt was established \textit{de facto}. 
After 30 years of the introduction of the green belt, the structure lost part of its multifractal signature, while after 60 years we can practically characterise the city as a monofractal structure. \\

In order to investigate the impact of such a natural barrier on the internal morphology of  the city, we performed a simulated experiment. We imposed a non-permeable barrier on a well-known multifractal diffusion process (DLA) and measured the evolution of its multifractal properties as the structure grew and began to approach the barrier. 
We obtained a clear transition from multifractal to monofractal behaviour in the two final stages of the  growth process of our structure. 
This is in very good correspondence with the observed behaviour for London.

The results presented here are hence another step towards the understanding of the implications of restricted urban growth policies in the urban morphology of modern cities. 

\appendix
\section{\label{sec:level7}Fractals review}

Monofractal objects are characterised by invariance under a change of length scale \cite{mandelbrot1977fractals, mandelbrot1983fractal}. When the change is performed in an isotropic way, the fractal is called \emph{self-similar}. If on the other hand, the length scale is changed in different directions through different factors, the fractal is called \emph{self-affine}.

The scaling exponent of the power-law relationship between the size of an object, e.g. its mass $M$, and the observable being measured, e.g. the length $L$, is in general identified as the fractal dimension $D$ \cite{havlin1991fractals}
\be
M \sim L^{D}.
\label{MLD}
\ee
In this sense, as the fractal grows, its density $\rho$ measured in the Euclidean space $d$ decreases:
\be
\rho(L) \sim L^{-\alpha}; \quad \alpha=d-D.
\label{rho_L} 
\ee
It is important to note, that this fractal dimension is not always unique. 
Different ways to characterise the system, and to measure scale-invariance with respect to different properties might lead to different exponents. 
For example, if a metric is defined in the space, one can find the scaling relationship between the number of units needed to cover the space as the fractal object is re-sized. 
If those units are balls, the exponent in this case corresponds to the \emph{Hausdorff-Besicovitch} dimension $D_H$ \cite{feder1988fractals,mandelbrot1985self}. This reduces to a problem of optimization of the number of balls needed to cover the space. 
In the case of simple self-affine fractals, the different fractal dimensions converge to a single value, and the fractal can be fully characterised by a single dimension $D=D_H$.
Nevertheless, this is seldom the case for fractals encountered in nature.

\subsubsection{Random Fractals}
Fractals are classified in terms of deterministic and random fractals \cite{vicsek1992fractal}. 
Deterministic fractals correspond to mathematically constructed objects whose scale invariance holds for all scales.
Examples of these are the well-known Sierpinski gasket and the Koch curve.
In experiments and in nature only random fractals are observed. These are finite objects that can be characterised as fractals within a specific regime of length scales. We denote by $a$ the smallest linear size that can be measured, and by $L_m$ the largest size, that might correspond to the size of the box where the finite fractal can be embedded in a Euclidean $d$-dimensional space \cite{havlin1991fractals}. 
For random fractals Eq.~(\ref{rho_L}) becomes [Meakin1990]:
\be
\rho(\lambda) \sim \lambda^{-\alpha}; \quad \lambda=\frac{L_m}{a}.
\label{rho_lambda} 
\ee 
In general the linear size $L$ should be replaced by $L/a$.
At this point it is very important to note, that random fractals exhibit self-affinity or self-similarity symmetries at a statistical level only.
This means that the correct way to measure the scaling exponent is through a large number of samples. 
The density ceases to be global, and one needs to take into account the average over all origins, and for non-isotropic fractals, over all orientations and ensemble realisations.  
The density is hence replaced by a density-density correlation function 
\be
c(\boldsymbol{r})=\frac{1}{V}\sum_{\boldsymbol{r}'}\rho(\boldsymbol{r}+\boldsymbol{r}')\rho(\boldsymbol{r}').
\label{densdens}
\ee
that can be seen as the probability of finding a particle at $\boldsymbol{r}+\boldsymbol{r}'$ if there is already one at $\boldsymbol{r}'$, and where the volume $V$ can be considered in terms of the total number of particles. 
For isotropic fractals, the density-density correlation function corresponds to the density distribution around $r$, since $c(\boldsymbol{r})=c(\boldsymbol{r}')$. This is scale-invariant 
\be
c(\eta r)\sim \eta^{-\alpha}c(r).
\label{self_simEq}
\ee
where $\eta$ denotes the re-scaling factor, and can also be expressed as
\be
c(r)\sim r^{-\alpha} .
\label{cralpha}
\ee
For growing systems whose mass can be characterised by the number of points (or particles), the fractal dimension can be obtained by looking at the scaling relationship between the number of points $M(R)$ within a sphere of radius $R$ \cite{vicsek1990mass}
\be
M(R)\sim \int_0^R c(r) d^dr \sim R^{d-\alpha}; \quad D=d-\alpha.
\label{massradiusEq}
\ee
Note that this equation only holds at the asymptotic limit, i.e. $L_m$ and $R \gg a$.
In addition, this radius can be seen as the \emph{radius of gyration}, and the system can be considered as a growing system as presented above, or as a system that contains several fractals of different sizes that are the outcome of the same process. 
For this latter case, the average needs to be considered, and in order to reduce statistical fluctuations, a large number of samples is required.
In a different paper we will investigate this scenario, considering cities in a country as the different realisations of the same process.
The observed fluctuations are the result of the statistical nature of the self-similarity of these systems. 

Obtaining a good estimate for the fractal dimension is not an easy task \cite{falconer1997techniques}. 
For small scales of $L$, corrections to the above given equations need to be introduced.
In addition, further inaccuracies arise from the limitations of the sample size and the orders of magnitude involved in the distribution of fractals from which $D$ is estimated. 

One solution is to verify that all such fractals involved in the computation of $D$ are self-similar. 
This is done by collapsing all the curves for the different fractals to a single one using the equation for self-similarity Eq.~(\ref{self_simEq}) and by taking the characteristic length of the system as the scaling factor 
\be
\eta=\frac{1}{R}\sim M^{-1/D}.
\ee
leading to
\beal
c(r/R) &\sim \Big( \frac{1}{R}\Big)^{-\alpha}c(r),\\
c(r) &\sim R^{-\alpha}f(r/R),\\
 &\sim M^{(D-d)/D}f(r/{M^{1/D}}).
\end{align}

\subsubsection{Hurst exponent}

The fluctuations of a system around a specific pattern can also be characterised in random fractals that are self-affine through the \emph{Hurst exponent} $H$ \cite{vicsek1990multifractal,meakin1998fractals}.
Self-affine fractals present symmetries that are not isotropic, leading to different scaling factors in different directions. In a 2-d space, the coordinates in the $x$ and $y$ axes might be re-scaled via $\eta_x$ and $\eta_y$, such that: $\eta_y=\eta_x^{H}$.
The above-mentioned fluctuations can also be observed in time series, and $H$ will give an indication of the randomness of the data, or in single valued functions. 
If time for example is re-scaled by $b$ (for $b>0$), the whole function $F(t)$ needs to be re-scaled by a different factor 
\be
F(t)\simeq b^{-H}F(bt).
\ee
An important example is the Brownian motion of a particle [Meakin1990]. Let us denote by $B(t)$ the distance covered by a particle in time $t$. 
In this case, if $t\rightarrow bt$, then the process $B(bt)$ is invariant if and only if a change in the distance scale of $b^{1/2}$ is introduced simultaneously: $b^{1/2}B(t)\simeq B(bt)$.  
Hence $H=1/2$ in this case, confirming self-affine symmetry.

Other important examples are the financial time series. There is a lot of controversy with respect to the scaling properties of such time series \cite{lebaron2001stochastic,mandelbrot2001stochastic,gabaix2003theory} and there are views that such processes should be considered as multi-scaling processes, see \cite{di2007multi} for a review.

\section{\label{sssec:level2} Practical calculation of the multifractal measures}
The  formalism introduced in Sec.\ref{ssec:level2} presents several practical disadvantages. The two most important ones are: 1) the smoothing of the curve $D_q$, and 2) the numerical calculations of the derivative in (\ref{alpha1}) and (\ref{q_d}). To overcome these difficulties, we apply the method described in \cite{chhabra1989direct} to obtain the parameters $D_q$ , $\tau(q)$,  $\alpha(q)$ and $f(\alpha(q))$.

We proceed as follows:  
\begin{enumerate}{}{}
\item Calculate the number of non-empty boxes of size $\epsilon, N(\epsilon)$, necessary to cover the SIPP. 
\item Calculate the number of intersections at the box \textit{i} of size $\epsilon, \mu(i,\epsilon)$, which is the discrete version of Eq.~(1).
\item Calculate $P(i,\epsilon)$ given in Eq.~(4) 
\item Calculate the partition function of Eq.~(6) :
\begin{equation}
Z(q,i,\epsilon)=\sum{^{N(\epsilon)}_{i=1}}P(i,\epsilon)^q.
 \label{13}
\end{equation}
\item Apply the normalised measure $\mu(q)$
\begin{equation}
\mu(q,i,\epsilon)=\frac{P(i,\epsilon)^q}{Z(q,i,\epsilon)}.
 \label{14}
\end{equation}
\item Obtain by linear regression, $\alpha(q,\epsilon)$, $f(q,\alpha,\epsilon)$ and $\tau(q,\epsilon)$ :
\begin{equation}
 \alpha(q,\epsilon)=\Sigma{^{N(\epsilon)}_{i=1}}\mu(q,i,\epsilon)\ln(P(i,\epsilon)).
 \label{15}
\end{equation}
\begin{equation}
 f(q,\alpha,\epsilon) = \Sigma{^{N(\epsilon)}_{i=1}}\mu(q,i,\epsilon)\ln(\mu(q,i,\epsilon))).
  \label{16}
\end{equation}
\begin{equation}
 \tau(q,\epsilon)=\Sigma{^{N(\epsilon)}_{i=1}}P(i,\epsilon)^{q-1}.
 \label{17}
\end{equation}
\item Finally calculate $D_q$ according to
\begin{equation}
D(q,\epsilon) = \frac{\tau(q,\epsilon)}{q-1}.
 \label{18}
\end{equation}
\end{enumerate}

In practice, $q$ does not take values in the entire range $(-\infty , \infty)$, due, first, to the obvious computational limitations, second, to the inherent statistical errors associated with the linear regressions and third, to the specific dimensionality constraints of our SIPP.  
The topological representation of these networks is a two-dimensional point set and Eq.~(\ref{q_d}) implies that $f(\alpha(q)) \leq D,$ where D is the dimension of the substrate on which the measure is distributed (in our case $D=2$). But Eq.~(\ref{eqDq_lim}) can indeed generate values $D_q\geq D$ and $f(\alpha(q))\geq D$ for $q<0$.  In this research, initially we selected $q$ in the discrete interval $[-20,20]$ with a 0.25 step separation between each values, and the actual valid range is selected in two steps:
\begin{description}
\item[a] All the q values where $D_q$ and $f(\alpha(q))$ are greater than 2 are dismissed for the further analysis and,
\item[b] For the values obtained through a regression, we selected the ones such that $R^2>0.9$.
\end{description}

\begin{acknowledgments}
The authors thank Dr. Kiril Stanilov for providing the unique historical data set analysed in this work and for his valuable comments about the probable effects of the Green Belt over the GLA area.\\
RM, APM, EA and MB were partially funded by the European Research Council (ERC) MECHANICITY Project (249393 ERC-2009-AdG)

\end{acknowledgments}


\bibliography{manuscript}

\end{document}